\documentclass{article}

\usepackage{graphicx} 
\usepackage{amsmath}
\usepackage{amssymb}

\newcommand{\nth}[1]{$#1^{\mathrm{th}}$} 
\newcommand{\ud}{\mathrm{d}}
\newcommand{\at}{{\char '100}}
   
\newcommand{\DEF}{\stackrel{\mathrm{def}}{=}}
\newcommand{\DEFt}{\stackrel{\mbox{\rm\tiny def}}{=}}  
  
\newcommand{\sign}{\mathrm{sign}}

\newcommand{\ket}[1]{|#1\rangle} 
\newcommand{\kete}[1]{|\kern.3ex#1\kern.3ex\rangle}
\newcommand{\bra}[1]{\langle #1 |}
\newcommand{\brae}[1]{\langle\kern.3ex #1 \kern.3ex|}

\newcommand{\vecb}[1]{\mathbf{#1}}

\begin{document}

\title{Bounding the ground-state energy of a many-body system with the differential method}
\author{Amaury Mouchet\thanks{mouchet\at phys.univ-tours.fr}}
\date{Laboratoire de Math\'ematiques
                            et de Physique Th\'eorique \\
                            \textsc{(cnrs umr 6083)},\\ Universit\'e
                            Fran\c{c}ois Rabelais
                            Ave\-nue Monge, Parc de Grandmont 37200
  Tours,
                            France.}

\maketitle

\begin{abstract} 
  This paper promotes the differential method as a  new fruitful
  strategy for estimating a ground-state energy of a many-body system.
  The case of an arbitrary number of attractive Coulombian particles
  is specifically studied and we make some favorable comparison of the
  differential method to the existing approaches that rely on variational
  principles. A bird's-eye view of the treatment of more general
  interactions is also given.
\end{abstract}

{PACS : 24.10.Cn,  12.39.Jh, 03.65.Db, 05.30.Jp}

\section{Introduction}

There is little need to stress the great importance of the 
ground-state in many domains of quantum physics.  Nevertheless, computing the
lowest energy of most systems cannot be done analytically and
approximations are required. Some techniques, being very general, have
been well-known for several decades. For instance the
variational methods (Rayleigh-Ritz) or the  perturbative series (Rayleigh-Schr\"odinger)
have their roots in the pre-quantum era; much later, numerical
algorithms (numerical diagonalizations as well as Monte-Carlo
computations), supported by the increasing power of computers, have
been able to provide a tremendous precision on the ground-state of a
large variety of very complex systems.  However, it is a much more
difficult task to rigorously estimate the discrepancies between the
exact ground-state energy~$E_0$ and the approximated one.  In
particular, though variational methods naturally provide upper bounds
on~$E_0$, obtaining lower estimates requires more sophisticated
techniques (for instance the Temple-like methods
\cite[\S~XIII.2]{Reed/Simon78a}), some of them being very
system-dependent (\textit{e.g.}  the moment method proposed in
\cite{Handy/Bessis85a} for rational-fraction potentials or the
Riccati-Pad\'e method proposed in \cite{Fernandez+89a} for
one-dimensional Schr\"odinger equations).

\textbf{The optimized variational methods.} For a many-body system
governed by pairwise interactions, an interesting strategy is to
approximate~$E_0$ from below in terms of the ground-states of the
two-body subsystems \cite{Fisher/Ruelle66a}.  Such an approach has
been successfully applied to Coulombian (bosonic and fermionic)
systems of charged particles \cite{Lenard/Dyson67a} or
self-gravitating bosons \cite{LevyLeblond69a,Basdevant+90a}.
Clever refinements have been proposed that provide some very accurate
lower bounds of $E_0$ for the three-body~\cite{Basdevant+93a} and the
four-body systems~\cite{Benslama+98a}. Though not easily generalizable
to an arbitrary number of particles, these last optimized variational
methods can be applied to interactions that are not necessarily
Coulombian and  may be relevant for quarks
models~\cite{Basdevant+90b} where some inequalities between baryon and
meson masses represent theoretical, numerical and experimental
substantial information \cite[and references therein]{Nussinov83a,Richard01a}. In practice, the optimized
variational methods allow to efficiently treat some models that have
simple scaling properties, for instance when the two-body interaction
can be described by a purely radial potential of the form~$v(r)\propto
\sign(\beta)r^\beta$.  The main reason relies in the fact that, except the
Coulombian ($\beta=-1$) and the harmonic ($\beta=2$) interactions, the
exact form of the ground-state energy of the two-body problem is not
known.  Yet, one can still take advantage of the power-law behavior of~$v$ to obtain worthwhile lower bounds for the \emph{ratio} between the
$N$-body and the~$2$-body ground-state energies.

\textbf{The differential method.} Besides the variational and
perturbative techniques that are mentioned above, there exists a third
very general method for approximating the ground-state energy of a
quantum system, namely the differential method (see \cite[and
references therein for a historical track]{Mouchet05a,Mouchet05b})
whose starting point is recalled in section \ref{sec:diffmethod}.  for
the sake of completeness.  As for the variational methods, the
differential method call on  a family of trial functions that
supposedly mimic the ground-state and that allow for the construction of a
function (the average of the Hamiltonian in the former case, the
so-called local energy in the latter case) whose absolute extrema within the
chosen trial family provide bounds on the exact ground-state
energy~$E_0$.  One of the main advantages of the differential method
over the other ones is that no integral is required and, then
it allows to work, even analytically, with rather complicated trial functions 
by encapsulating some rich structure of the potential.   
It is also worth mentioning that the same test
function leads to both upper and lower bounds on~$E_0$ and then the
estimates comes with a rigorous window.  Though
applicable to many models (to systems involving a magnetic field, to
discrete systems, to non-Schr\"odinger equations, etc.) the major
inconvenient of the differential method is that it requires, as a
crucial hypothesis, the exact eigenfunction to remain non-negative
in the configuration space (it must have no node). 
Therefore it excludes any ground-state
whose spatial wave-function is antisymmetric under some permutations
of its arguments. As far as fermionic systems are involved, the
differential method will concern only those whose ground-state
eigenfunction remains symmetric under permutations of the spatial
positions of the identical particles.

The aim of this paper is to apply the differential method specifically
to a system made of~$N$ non-relativistic particles of masses~$m_i$ in
a ~$\textsc{d}$-dimensional space whose Hamiltonian has the form
\begin{equation}\label{eq:HNbody}
  \tilde{H}=\sum_{i=0}^{N-1}\frac{\vecb{p}^2_i}{2m_i}+V(\vecb{r}_0,\dots,\vecb{r}_{N-1})\;.
\end{equation}
When the $N$ particles located at $\{\vecb{r}_i\}_{i=0,\dots,N-1}$
interact only through pairwise potentials~$v_{ij}=v_{ji}$\;, $V$ is
given by
 \begin{equation}\label{eq:VNbody}
  V=\sum^{N-1}_{\substack{i,j=0\\i<j}}v_{ij}(\vecb{r}_{ij})
\end{equation}
where $\vecb{r}_{ij}\DEF\vecb{r}_j-\vecb{r}_i$.  The spin-dependent
interactions, if any, are assumed to be included somehow 
 in the scalar potential~$V$ and
$\tilde{H}$ will be supposed to act on spatial wave-functions only; in
other words the possible spin configuration have been factorized out
in one way or another.  When~$N=3$ and~$N=4$ and for power-law
potentials, this is the same kind of
systems to which the optimized variational method applies also.  We
will consider the Coulombian case in section~\ref{sec:coulomb} and
systematically  compare the estimates given by 
the variational methods and the differential method. 
 In section ~\ref{sec:generalvij},
general interactions are considered (not necessarily
power-law $v_{ij}$'s). This is of course relevant for estimating the
ground-state energy of a system where the spin-independent strong
interactions are dominant; for heavy enough quarks for instance, it is
known \cite{Jaczko/Durand98a} that the non-relativistic
form~\eqref{eq:HNbody} may be pertinent\footnote{Possible relativistic
  corrections may be
  included (for instance by considering the spinless Salpeter
  equation) since the differential method does not require a quadratic
  kinetic energy.}. At atomic scales, the method could be applied to
clouds made of neutral atoms where short-range interactions govern the
dynamical properties.

\section{The differential method}\label{sec:diffmethod}

\subsection{The general strategy}\label{subsec:generalstrategy}

The necessary but sufficient condition for the differential method to
work is the following: the Hamiltonian~$H$ has one bound state
~$\ket{\Phi_0}$, associated with energy~$E_0$, such that
$\Phi_0(q)\DEFt\langle q\ket{\Phi_0}$ remains non-negative in an
appropriate $q$-representation, say of spatial positions. For
a~$N$-body system governed by the Hamiltonian~\eqref{eq:HNbody}, the
dynamics in the center-of-mass frame corresponds to a reduced
Hamiltonian~$H$ whose ground-state\footnote{\label{fn:trap}We will
only consider the cases where at least one bound state exists.
Physically, this can be achieved with a confining external potential
(a ``trap'' is currently used in experiments involving cold
atoms). Formally, this can be obtained in the limit of one mass, say
$m_0$, being much larger than the others. The external potential
appears to be the $v_{0i}$'s, created by such an infinitely massive
motionless device. It will trap the remaining~$N-1$ particles in some
bound
states if the $v_{0i}$'s increase sufficiently rapidly with
the $r_{0i}$'s.}  $\Phi_0$ has precisely this positivity property in
the whole configuration space $\mathcal{Q}_N$ of the
$(N-1)\textsc{d}$ relative coordinates
$q_N\DEFt(\vecb{r}_1-\vecb{r}_0,\dots,\vecb{r}_{N-1}-\vecb{r}_0)$. This is
the Krein-Rutman theorem (see \cite[\S XIII.12]{Reed/Simon78a}).  For
each state $\ket{\varphi}$, the hermiticity of $H$ implies the
identity $\bra{\Phi_0}(H-E_0)\ket{\varphi}=0$. If we choose
$\ket{\varphi}$ such that its representation $\varphi(q_N)$ is a
smooth normalizable real wave-function, we obtain
\begin{equation}\label{eq:integralidentity}
        \int_{\mathcal{Q}_N}\Phi^*_0(q_N)(H-E_0)\varphi(q_N)\,dq_N=0\;.
\end{equation}
Taking into account the positivity of~$\Phi_0$ on $\mathcal{Q}_N$,
there necessarily exists some $q_N$ such that
$(H-E_0)\varphi(q_N)\geqslant0$ and some other configurations for
which $(H-E_0)\varphi(q_N)\leqslant0$.  Choosing $\varphi>0$ on
$\mathcal{Q}_N$, we get both an upper and a lower bound on $E_0$:
\begin{equation}\label{eq:inequalities}
        \inf_{\mathcal{Q}_N} \big(E_\mathrm{loc}^{[\varphi]}(q_N)\big)
        \leqslant\
        E_0\ \leqslant\sup_{\mathcal{Q}_N} \big(E_\mathrm{loc}^{[\varphi]}(q_N)\big)\;,
\end{equation}
 where the local energy is defined by
\begin{equation}
        \label{def:localenergy}
        E_\mathrm{loc}^{[\varphi]}(q_N)\ \DEF\ \frac{H\varphi(q_N)}{\varphi(q_N)}\;.
\end{equation}
In other words, the differential method provides an estimate
\begin{equation}\label{eq:Edm}
  E_0^{\mathrm{(d.m.)}}\ \DEF\ \frac{1}{2}\left[
\sup_{\mathcal{Q}_N} \big(E_\mathrm{loc}^{[\varphi]}(q_N)\big)
+ \inf_{\mathcal{Q}_N} \big(E_\mathrm{loc}^{[\varphi]}(q_N)\big)\right]
\end{equation}
that comes with a rigorous windows $\pm\Delta E_0^{\mathrm{(d.m.)}}$ where
\begin{equation}\label{eq:deltaEdm}
  \Delta E_0^{\mathrm{(d.m.)}}\ \DEF\ \frac{1}{2}\left[
\sup_{\mathcal{Q}_N} \big(E_\mathrm{loc}^{[\varphi]}(q_N)\big)
- \inf_{\mathcal{Q}_N} \big(E_\mathrm{loc}^{[\varphi]}(q_N)\big)\right]\;.
\end{equation}

Unlike for the variational method, the determination of the absolute
extrema of the local energy does not require the computation of any
integral. Even the norm of the test function $\varphi$ is not required
provided it remains finite.  The two
inequalities~\eqref{eq:inequalities} become equalities (the local
energy becomes a flat function) when $\varphi=\Phi_0$ and therefore we
will try to construct a test function that mimics $\Phi_0$ at best. We
will choose $\varphi$ that respects the \textit{a priori} known
properties of $\Phi_0$: its positivity, its boundary conditions and
its symmetries if there are any. Since for each test function the
error on $E_0$ is controlled by inequalities~\eqref{eq:inequalities},
the strategy for obtaining decent approximations is clear: First,
we must choose or construct $\varphi$ to eliminate all the singularities
of the local energy in order to work with a bounded function.
For instance, when the Hamiltonian has the form $p^2+V$
with  $V$ being  unbounded at some finite or infinite distances,
 the first kinetic term 
of the local energy $E_\mathrm{loc}^{[\varphi]}=-\Delta\varphi/\varphi+V$ 
must compensate the singular behavior of $V$ for the corresponding configurations (we will
work systematically with units such that~$\hbar=1$).
Once a bounded local energy, say $E_\mathrm{loc}^{[\varphi_0]}$, is obtained, we can proceed 
to a second step: perturb the test function, $\varphi_0\to\varphi=\varphi_0+\delta\varphi$,
 in the neighborhood of the absolute
minimum (resp. maximum) of $E_\mathrm{loc}^{[\varphi_0]}$ in order to 
increase~$\min E_\mathrm{loc}^{[\varphi_0]}$
(resp. decrease~$\max E_\mathrm{loc}^{[\varphi_0]}$). Up to the end of this article, we
will focus on the first step: we will show how obtaining a bounded
local energy furnishes some sufficiently constrained guidelines for
obtaining reasonable bounds on $E_0$\footnote{One can understand it
from the extreme sensitivity of the local energy to any local
perturbation of the test function: while, in the variational methods,
the quantity $\bra{\varphi}H\ket{\varphi}/\langle\varphi\ket{\varphi}$
is quite robust to local perturbations because it represents precisely
an average on the configuration space, the local energy may become
unbounded quite easily by canceling locally $\varphi$ faster than
$H\varphi$.}.  We will keep for future work the systematic local
improvements of the absolute extrema of the local energy. In
\cite[\S~6]{Mouchet05a}, I have shown on a simple example how this can
be done.

\subsection{Illustration in the two-body case}

Before coping with much complex systems, let us first consider the
case of the two-body problem that can be reducible to a
one single non-relativistic particle of unit mass in an
external potential $V$. The local energy is
\begin{equation}
  E_\mathrm{loc}^{[\varphi]}
=-\frac{\Delta\varphi}{2\varphi}+V=-\frac{1}{2}\Delta S-\frac{1}{2}(\nabla S)^2
+V
\end{equation}
where $S\DEF\ln(\varphi)$ is a well-defined function when~$\varphi>0$. 

For the $\textsc{d}$ Coulombian potential $V(r)=\kappa/r$
($\textsc{d}>1$), the singularity at $r=0$ controls the local
behavior of the test function if one wants a bounded local energy. If
$\lim_{r\mathop{\to}0}S(r)$ is finite, by possibly subtracting an
irrelevant constant term, we can suppose that this limit vanishes.
Therefore, without too much loss of generality, we assume that $S$,
can be asymptotically expanded near~$r=0$ on a family of power
functions (for $\textsc{d}=1$, the logarithmic functions should be
considered) whose dominant term can be written
like~$S(r)\sim_{r\mathop{\to}0} s_0 r^{\sigma+1}/(\sigma+1)$
with~$\sigma\neq-1$ and~$s_0\neq0$.  Balancing the dominant terms in
the local energy, it is therefore straightforward, to check that the
only choice for the parameters~$s_0$ and~$\sigma$ to get rid of the
Coulombian singularity is too take $\sigma=0$
and~$s_0=2\kappa/(\textsc{d}-1)$.  It happens that for~$S$ exactly equal
to $2\kappa r/(\textsc{d}-1)$, we obtain a global constant local
energy, namely $-2\kappa^2/(\textsc{d}-1)^2$. Therefore we
have obtained the exact wave-function of the ground-state provided we
eventually check that the wave-function is square-integrable which is true
 for~$\kappa<0$.

For the harmonic oscillator $V(r)=\omega^2 r^2/2$, $V$ is unbounded
as~$r$ increases ($r=+\infty$ is a singular point for~$V$). If we
tentatively look for an $S$ whose asymptotic expansion at
$r\to+\infty$ has a leading term of the form $s_0 r^{\sigma+1}/(\sigma+1)
$, we necessarily get $\sigma=1$ and $s_0=-\omega$ (the
cases where $\sigma\leqslant-1$ are ruled out by the square-integrable
property). The local energy $\textsc{d}\omega/2$ is actually
constant for all~$r$'s and indeed represents the exact ground-state
energy. More generally, with the help of standard  linear algebra arguments,
 for $V$ being any definite positive quadratic 
form, we can always find a quadratic form $S$ for which 
the local energy is globally constant.

\subsection{Formulation for the many-body problem}

When the potential~$V$ has the form \eqref{eq:VNbody}, a natural
choice of trial function is to take (for variational techniques in
a few nuclear body context,
such a choice has been used by \cite{Bodmer/ShamsherAli64a,VanDyke/Folk69a} for instance)
\begin{equation}\label{eq:totalphi}
  \varphi(q_N)=\prod^{N-1}_{\substack{i,j=0\\i<j}}\phi_{ij}(\vecb{r}_{ij})
\end{equation}
where each of the $N(N-1)/2$ functions
$\phi_{ij}(\vecb{r})=\phi_{ji}(-\vecb{r})$ depends on~$\textsc{d}$
coordinates\footnote{The present paper wants mainly to stress the
simplicity of the differential method. It does not seek for a real
performance at the moment and we will not try to improve the choice of
coordinates. Working with Jacobi coordinates, for instance, or
constructing optimized coordinates as done in~\cite{Benslama+98a} may
lead to better results. Anyway, we will see that
 the numerical results of section~\ref{sec:coulomb}
 are satisfactory enough for validating the approach
by the differential method.}.

It is straightforward to check that this choice describes a state with
a fixed center-of-mass: indeed we have
$(\sum_{i=0}^{N-1}\vecb{p}_i)\ket{\varphi}=\vecb{0}$. Hence,
$\tilde{H}\varphi=H\varphi$ and the local energy is given by
\begin{equation}\begin{split}
  E_\mathrm{loc}^{[\varphi]}(q_N)=
  \sum^{N-1}_{\substack{i,j=0\\i<j}}
    \left(
      -\frac{1}{2m_{ij}}\frac{\Delta\phi_{ij}(\vecb{r}_{ij})}{\phi_{ij}(\vecb{r}_{ij})}+v_{ij}(\vecb{r}_{ij})
                         \right)\\
  -\sum_{\widehat{j,i,k}}\frac{1}{m_i}\,
   \frac{\boldsymbol{\nabla}\!\phi_{ij}(\vecb{r}_{ij})}{\phi_{ij}(\vecb{r}_{ij})}\,\cdot\,
   \frac{\boldsymbol{\nabla}\!\phi_{ik}(\vecb{r}_{ik})}{\phi_{ik}(\vecb{r}_{ik})}
\end{split}
\end{equation}
where $m_{ij}$ stands for the reduced masses $m_im_j/(m_i+m_j)$.  The
last sum involves all the $N(N-1)(N-2)/2$ angles $(\widehat{j,i,k})$
between~$\vecb{r}_{ij}$ and~$\vecb{r}_{ik}$
that can be formed with all the triangles made of three particles
having three distinct labels ($i\neq j$, $i\neq k$, $j\neq k$).  Let
us now take~$\phi_{ij}$ to be a positive solution of the two-body
spectral equation
 \begin{equation}\label{eq:schro2body}
   -\frac{1}{2m_{ij}}\Delta\phi_{ij}+v_{ij}\phi_{ij}=\epsilon_{ij}\phi_{ij}\;.
\end{equation}           
The local energy becomes
\begin{equation}\label{eq:NbodyElocSS}
   E_\mathrm{loc}^{[\varphi]}(q_N)= \sum^{N-1}_{\substack{i,j=0\\i<j}}\epsilon_{ij}
   -\sum_{\widehat{j,i,k}}\frac{1}{m_i}\,
    \boldsymbol{\nabla}\!S_{ij}(\vecb{r}_{ij})\cdot
    \boldsymbol{\nabla}\!S_{ik}(\vecb{r}_{ik})
\end{equation}
where $S_{ij}\DEF\ln(\phi_{ij})$. The trial
wave-function~\eqref{eq:totalphi} of the global system must be kept
square-integrable but it is not necessary for \emph{all} two-body
subsystems to have a bound state when isolated\footnote{But for some
  pairing, \eqref{eq:schro2body}
 must have a normalizable solution. The cases of
  Borromean states where no two-body binding is possible\cite{Richard03a}
cannot be described by the form~\eqref{eq:totalphi} if we keep \eqref{eq:schro2body}.}. For instance,
 for two electric charges having the same sign a
positive but non-normalizable solution of \eqref{eq:schro2body} can be
found.  When~$v_{ij}$ admits at least one bound state (see also
footnote~\ref{fn:trap}), thanks to the Krein-Rutman theorem, we are
certain to get a positive~$\phi_{ij}$ when taking the ground-state of
the two-body system and $\epsilon_{ij}$ its corresponding energy.  At finite distances, if
the possible singularities of $v_{ij}$ are not too strong, we
expect that~$\phi_{ij}$ and then~$S_{ij}$ to be smooth enough
for~$E_\mathrm{loc}$ to remain bounded. At infinite distances,
$E_\mathrm{loc}$ is expected to become infinite if $v_{ij}$ does not
tend to a constant sufficiently quickly.  To see that, one can take
purely radial potentials.  \textit{i.e} $v_{ij}(\vecb{r})=v_{ij}(r)$
where $r\DEFt||\vecb{r}||$, and consider the asymptotic behavior of
$S_{ij}$ given by the semiclassical (\textsc{jwkb}) theory (see for
instance \cite{Maslov/Fedoriuk81a}). Its derivative is given
by
$S'_{ij}(r)\sim_{r\to\infty}-\sqrt{2m_{ij}[v_{ij}(r)-\epsilon_{ij}]}$
and is not 
bounded
 if $v_{ij}$ is not (at infinite distances). Strictly speaking, it is
only for short-distant potentials that we can hopefully obtain
rigorous non-trivial inequalities~\eqref{eq:inequalities} while keeping the
choice \eqref{eq:totalphi} with \eqref{eq:schro2body}. However,
as will be discussed  in section~\eqref{sec:generalvij},
the ground-state 
 energy may be generally not be very sensitive
to the potential at large distances (far away where $\Phi_0$ is
localized) and this physical assumption may be implemented
by introducing a cut-off length from the beginning.

  For purely radial
potentials \eqref{eq:NbodyElocSS} simplifies in
\begin{equation}\label{eq:NbodyElocSSradial}
   E_\mathrm{loc}^{[\varphi]}(q_N)= \sum^{N-1}_{\substack{i,j=0\\i<j}}\epsilon_{ij}
   -\sum_{\widehat{j,i,k}}\frac{1}{m_i}\, S'_{ij}(r_{ij})S'_{ik}(r_{ik})\cos(\widehat{j,i,k})\;.
\end{equation}

Yet, for a multidimensional, non-separable, Schr\"odinger equation
like \eqref{eq:schro2body}, a \textsc{jwkb}-like asymptotic expression is
generally not available
\cite[Introduction]{Maslov/Fedoriuk81a}. Nevertheless, the
differential method is less demanding than the semiclassical
approximations: we will try to keep the local energy, like the one
given by~\eqref{eq:NbodyElocSS}, bounded at infinity but we will not
necessarily require it to tend to the \textit{same} limit in all
directions.

\section{The Coulombian  problem}\label{sec:coulomb}

The purely Coulombian problem in $\textsc{d}>1$ dimensions corresponds to the situation where 
 all $v_{ij}$'s are radial potentials and have the form
\begin{equation}
 v_{ij}(r)=\frac{e_{ij}}{r} 
\end{equation}
for $N(N-1)/2$ coupling constants~$e_{ij}$ that may be or may be not
constructed from individual quantities like charges.  
Provided a $N$-body ground-state exists, we can solve exactly \eqref{eq:schro2body}
making use of~ $\Delta\phi(r)=\phi''(r)+(\textsc{d}-1)\phi'(r)/r$. We obtain a
bounded local energy given by
\begin{equation}\label{eq:ElocNbodyCoulomb}
   E_\mathrm{loc}^{[\varphi]}(q_N)=\sum^{N-1}_{\substack{i,j=0\\i<j}}-\frac{2m_{ij}e_{ij}^2}{(\textsc{d}-1)^2}-
\frac{4}{(\textsc{d}-1)^2}\sum_{(\widehat{j,i,k})}\frac{m_{ij}m_{ik}e_{ij}e_{ik}}{m_i}\cos(\widehat{j,i,k})\;.
\end{equation}
For obtaining upper and lower bounds on~$E_0$, one has just to
calculate the absolute extrema of such a function.  It can be done by
standard optimization routines up to quite large~$N$ and even
analytically in some cases (see below). The recipe is therefore simple
and systematic: as far as only Coulombian interactions are involved,
we can work with generic masses and coupling constants for
which~\eqref{eq:totalphi} is normalizable.  The remaining of this
section will concern the quality of these bounds and then we will
accord our attention to cases that have been treated by other methods,
mainly those treated in the references cited in the second paragraph
of the introduction.  More specifically, in order to leave aside the
problem of the existence of a ground-state we will consider the
case of attractive interactions 
only\footnote{For
an immediate application in the case of charged electric particles see
\cite{Mouchet05a} where the Helium atom is discussed. While the differential method provides
an analytical non-trivial upper bound
for the ground-state of the Helium-like atoms, $E_0\leqslant-(Z-1/2)^2$ 
in the simplest model ($\textsc{d}=3$, non-relativistic, spinless and with an infinitely massive nucleus
of charge $Z$ in atomic units),
in the case of more delicate systems like the positronium ion ($e^+,e^-,e^-$),
a systematic improvement is clearly required but is beyond the scope of this paper
as explained at the end of subsection~\ref{subsec:generalstrategy}. Indeed, for ($e^+,e^-,e^-$)
if we content ourselves with eliminating the singularities, we obtain, for $\textsc{d}=3$,
$-9m\alpha^2/8\leqslant E_0\leqslant0$ ($\alpha$ being the fine structure constant).
 The upper bound is trivial while the lower bound is even worse
compared to $-3m\alpha^2/4$ obtained in \cite[eq. (6.4)]{Basdevant+90b}
or to $-3m\alpha^2/4$  obtained  by a simple crude argument \cite[eq. (6.6)]{Basdevant+90b}
(the exact result is $E_0\lesssim-m\alpha^2/4$).} (all $e_{ij}$'s being negative).

\subsection{Arbitrary number of identical attractive particles}\label{subsec:Nidentical}

In this section we consider one species of particles only: for all $i$
and $j$ we denote~$m_i=m$ and $e_{ij}=-g^2$. The local
energy~\eqref{eq:ElocNbodyCoulomb} becomes
\begin{equation}\label{eq:EFN}
  E_\varphi(q_N)=-\frac{\epsilon_0}{(\textsc{d}-1)^2}\left(\frac{1}{2}N(N-1)+F_N(q_N)\right)
\end{equation}
where $\epsilon_0\DEF mg^2$. The function
\begin{equation}\label{def:FN}
   F_N(q_N)\DEF\sum_{(\widehat{j,i,k})}\cos(\widehat{j,i,k})
\end{equation}
is invariant under translations and rotations but also under
dilations of the particle configuration.  For~$N=3$, the
appendix~\ref{sec:appendixA} proofs that $\sup_{\mathcal{Q}_3}F_3=3/2$
is reached when the three particles make an equilateral triangle and
$\inf_{\mathcal{Q}_3}F_3=1$ is obtained when they are aligned.  From
this last result we are able to provide the lower bounds for~$F_N$ for
any~$N$: by a decomposition of~$F_N$ into a sum on~$N(N-1)(N-2)/6$
triangle contributions,
 \begin{equation}
 F_N(q_N)=\sum_{\substack{\{i_1,i_2,i_3\}\\ 1\leqslant i_1<i_2<i_3\leqslant N}}
\underbrace{\cos(\widehat{i_3,i_1,i_2})+\cos(\widehat{i_1,i_2,i_3})+\cos(\widehat{i_2,i_3,i_1})}_
{\displaystyle =F_3(\vecb{r}_{i_1},\vecb{r}_{i_2},\vecb{r}_{i_3})}\;,
\end{equation}
 all the
$F_3$'s in the sum reach their minimum simultaneously when all the particles are aligned,
 and for this
configuration we
have~$\inf_{\mathcal{Q}_N}F_N=N(N-1)(N-2)/6$. From~\eqref{eq:EFN}, we
deduce that for each~$N$
\begin{equation}
  E_0\leqslant-\frac{\epsilon_0}{6(\textsc{d}-1)^2}\,N(N-1)(N+1)\;.
\end{equation} 
For~$\textsc{d}=3$, the same exponential test-functions lead to the better
variational estimate \cite[eq.~(17)]{LevyLeblond69a}
\begin{equation}\label{eq:upperboundJMLL}
  E_0\leqslant-\epsilon_0\frac{25}{512}\,N(N-1)^2
\end{equation}
($25/512\simeq0.0488\gtrsim1/24\simeq0.0417$).  This was expected from
the general identity valid for any normalized function~$\varphi$,
\begin{equation}
  \int_\mathcal{Q}\varphi^*(q)H\varphi(q)\,dq
=\int_\mathcal{Q}|\varphi(q)|^2 E_\mathrm{loc}^{[\varphi]}(q)\,dq
\leqslant \sup_\mathcal{Q} \big(E_\mathrm{loc}^{[\varphi]}(q)\big)\;.
\end{equation}
The differential method always gives worse upper bounds than the
variational method with the same test-functions but, in the last case,
one still has to be able to compute the integrals and one cannot
generally estimate how far from the exact value the average Hamiltonian
is.  For a different choice of test functions, a better variational
upper-bound has been obtained~\cite[eq. (16)]{Basdevant+90a},
\begin{equation}\label{eq:upperboundBasdevant}
  E_0<-.0542N(N-1)^2\;.
\end{equation}
 
As far as lower estimates are concerned, bounding $F_N$ from above
will allow us to improve the existing results, namely (for $\textsc{d}=3$)
\begin{equation}\label{eq:minorationBasdevant}
  E_0\geqslant-\frac{1}{16}N^2(N-1)
\end{equation}
obtained in \cite[eq. (12)]{Basdevant+90a}. 
 
\begin{figure}[!ht]
\center
\includegraphics[width=12cm]{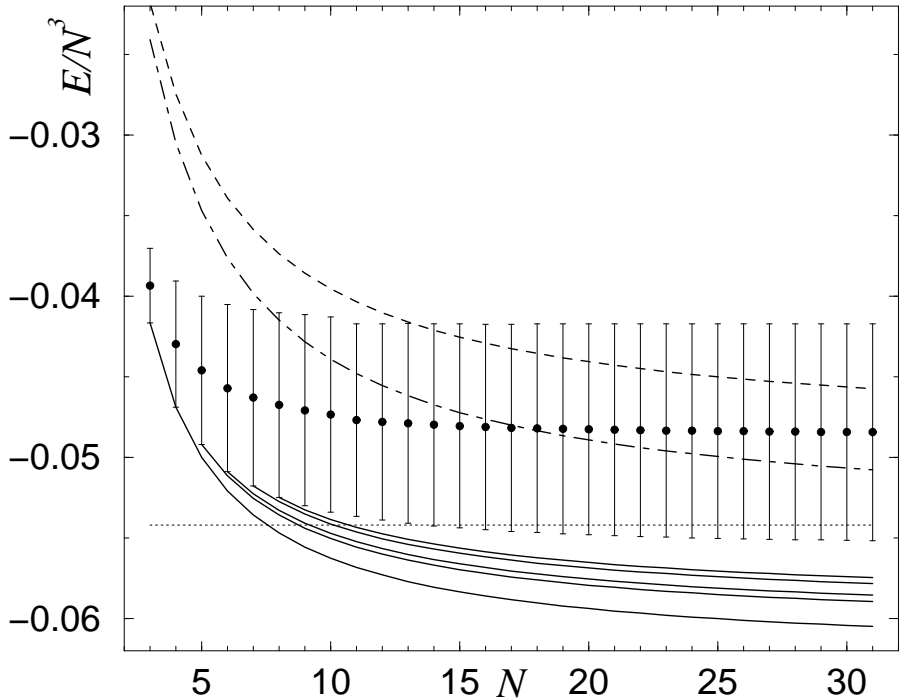}
\caption{\label{fig:Emd_Ncoulombid}
Estimations of ground-state energy for $N$ identical attractive
Coulombian particles (rescaled by a factor~$N^3$). The differential method 
provides~$E_0^{\mathrm{(d.m.)}}$ defined by~\eqref{eq:Edm} (black disks) with exact error bars
whose half-width are given by~\eqref{eq:deltaEdm}. Each solid line starting at~$M$ corresponds
to the lower bounds~\eqref{eq:E0minalphaM}. For $M=3$ and $M=4$, there is only one solid curve
given by~\eqref{eq:minorationBasdevant}. The dashed line 
 corresponds to the upper bound
\eqref{eq:upperboundJMLL} given  by \protect\cite[eq.~(17)]{LevyLeblond69a}. 
 The dot-dashed line  corresponds to the upper bound
\eqref{eq:upperboundBasdevant} given by \protect\cite[eq. (16)]{Basdevant+90a}
that tends,  when $N\to\infty$, to the thin dotted
horizontal line located at $-.0542$. 
}
\end{figure}

First, when $N$ is not too large for the numerical computation to
remain tractable, the direct calculation of~$\sup_{\mathcal{Q}_N}F_N$
shows (see figure~\ref{fig:Emd_Ncoulombid}) that it gives better lower estimates
than~\eqref{eq:minorationBasdevant}.  For very large $N$, we can
nevertheless benefit from the maximum of $F_M$ for
smaller~$M$. Indeed, for $M\leqslant N$ we can decompose~$F_N$ into
contributions of~$M$-clusters as follows:
\begin{equation}\label{eq:clustering}
  F_N(q_N)=\sum_{M-\text{subclusters}}\frac{(M-3)!\,(N-M)!}{(N-3)!}\,F_M(q_M)
\end{equation}
where the sum is taken on all the $M$-subclusters, labeled by the
coordinates~$q_M$, that can be formed with the given
configuration~$q_N$.  This sum involves exactly $N!/M!/(N-M)!$ terms
and  we have
\begin{equation}\label{eq:alphaMalphaN}
  \sup_{\mathcal{Q}_N}F_N\leqslant\frac{N(N-1)(N-2)}{M(M-1)(M-2)}\;\sup_{\mathcal{Q}_M}F_M\;.
\end{equation}
This leads to define 
\begin{equation}\label{eq:alphaM}
\alpha_M\DEF\frac{\sup_{\mathcal{Q}_M}F_M}{M(M-1)(M-2)}
\end{equation}
and from \eqref{eq:EFN} we find
\begin{equation}\label{eq:E0minalphaM}
  E_0\geqslant-\frac{\epsilon_0}{(\textsc{d}-1)^2}\,N(N-1)\left(\frac{1}{2}+\alpha_M(N-2)\right)\;.
\end{equation}
Since, from \eqref{eq:alphaMalphaN}, $\alpha_M$ is decreasing when $M$
increases, the larger $M$ the better the lower estimate of $E_0$.

For~$M=3$ we have already seen that~$\alpha_3=1/4$; for $\textsc{d}=3$,
\eqref{eq:E0minalphaM} reproduces
exactly~\eqref{eq:minorationBasdevant}. No better estimate is obtained
when considering~$M=4$.  Indeed, the configuration of particles that
maximizes~$F_4$ corresponds to the regular tetrahedron because its
faces, that are equilateral triangles, maximize the contributions of
all the 3-subclusters simultaneously.  We obtain
$\sup_{\mathcal{Q}_4}F_4=6$ and hence $\alpha_4=\alpha_3$.

\begin{figure}[!ht]
\center
\includegraphics[width=10cm]{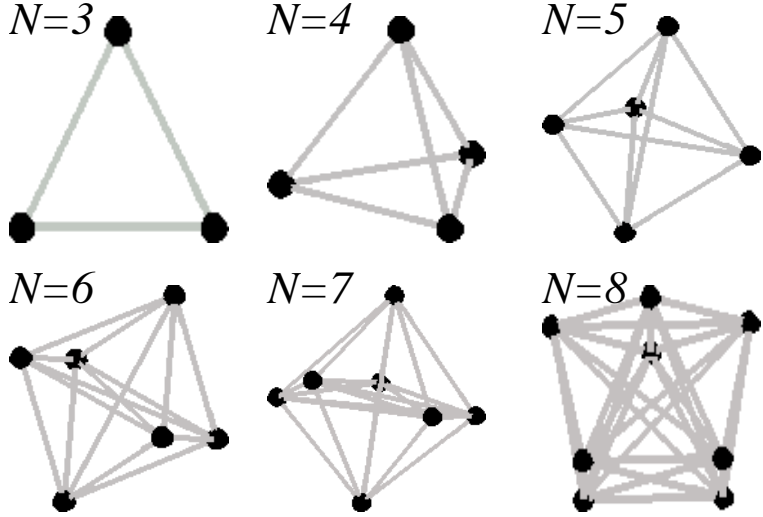}
\caption{\label{fig:Fmax}The configurations that maximize $F_N$ defined by~\eqref{def:FN} for
 identical Coulombian attractive particles.}
\end{figure}

For~$M=5, 6, 7, 8$ and $\textsc{d}=3$, the configurations that maximize~$F_M$
can be seen in figure~\ref{fig:Fmax}.  Crossed numerics and analytical studies lead to
 very plausible
conjectures on the geometrical description of the configuration
for~$M=5$ and~$M=8$ for which explicit analytical value of~$\alpha_M$
can be proposed~\cite{Mouchet05b}. The lower bound for
\eqref{eq:E0minalphaM}, for $N\geqslant M$, is strictly improved when
increasing~$M$ from~5 and in particular is better
than~\eqref{eq:minorationBasdevant}.
 
\begin{figure}[!Ht]
\center
\includegraphics[width=10cm]{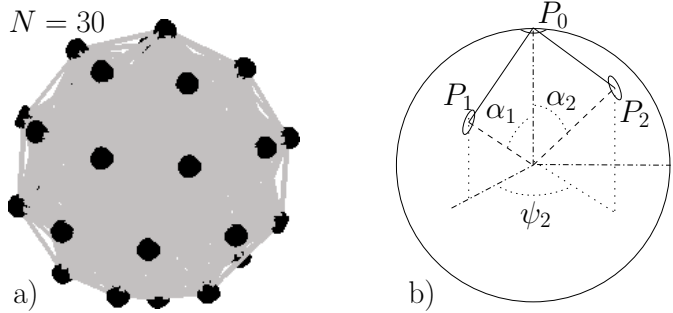}
\caption{\label{fig:Fmaxinfini}For large $N$, $F_N$ is maximized for a configuration where 
the $N$ points seems, numerically, to distribute uniformly on a sphere (with the well-know caveat
concerning the ambiguity of such a notion \protect\cite{Saff/Kuijlaars97a}). 
Such an optimal configuration for~$N=30$ is shown in a). The continuous limit of ~$F_N/N^3$
when~$N\to\infty$ 
can be computed with the help of figure b). }
\end{figure}

However the sequence of improvements obtained this way seems to
saturate up to~$\alpha_\infty=2/9$:
\begin{multline}
  \alpha_3=\alpha_4=\frac{1}{4}\geqslant\alpha_5\simeq.2432\geqslant\alpha_6\simeq.2414\geqslant\alpha_7\simeq.2382\geqslant
\alpha_8\simeq.2366\geqslant\cdots\\
\cdots\geqslant\alpha_{30}\simeq.2266\cdots\geqslant\alpha_\infty=\frac{2}{9}\simeq.2222\;.
\end{multline}
From the optimized configuration obtained with~$N$ about several tens
(see figure~\ref{fig:Fmaxinfini}a), we can hopefully guess that the limit~$N\to\infty$ leads
to a continuous and uniform distribution of the particles on the same
sphere.  The continuous limit of~$\sup_{\mathcal{Q}_N}F_N$ varies
as~$N^3$ with~$N$. If, on the unit sphere~$\mathcal{S}$, the~$N$
particles get distributed uniformly with density~$\sigma=N/4\pi$, the
continuous limit of $\sup_{\mathcal{Q}_N}F_N$ is given by ($\ud S_i$ is
an infinitesimal portion of the sphere near the point~$P_i$,
$i=0,1,2$; see figure~\ref{fig:Fmaxinfini}b)

\begin{eqnarray*}
  \sup_{\mathcal{Q}_N}F_N&\!\sim\!&
  \int_\mathcal{S}\sigma\, \ud S_0\;\frac{1}{2}\int_\mathcal{S}\sigma\, \ud S_1\int_\mathcal{S}\sigma\, \ud S_2\,
  \cos(\widehat{P_1,P_0,P_2})\;;\\
 &\!\sim\!&\frac{1}{2}\sigma^3\,8\pi^2\int_{0}^\pi \ud\alpha_1\int_{0}^\pi \ud\alpha_2
 \int_{0}^{2\pi} \ud\psi_2\,\sin(\alpha_1)\sin(\alpha_2)\cos(\widehat{P_1,P_0,P_2}).
\end{eqnarray*} 
When
$\cos(\widehat{P_1,P_0,P_2})=\frac{\mathbf{P_0P_1}}{||\mathbf{P_0P_1}||}\cdot\frac{\mathbf{P_0P_2}}{||\mathbf{P_0P_2}||}$
is expressed as a function of~$\alpha_1$, $\alpha_2$ and~$\psi_2$, we
straightforwardly
get~$\sup_{\mathcal{Q}_N}F_N/N^3\sim_{N\to\infty}2/9=\alpha_\infty$.
For infinite $N$, this last result supplement
 the upper bound given by~\eqref{eq:upperboundBasdevant} and we have
\begin{equation}
  N\to\infty;\qquad -\frac{1}{18}\simeq-.0556 \lesssim\
        \frac{E_0}{N^3}\ \lesssim-.0542\;.
\end{equation}

\subsection{Three particles with one different from the two others}\label{subsec:11m}
Without loss of generality, we can choose units
where~$e_{12}=e_{13}=e_{23}=-1$, ~$m_1=m_2=1$, $m_0=m$. For~$\textsc{d}=3$ the
local energy~\eqref{eq:ElocNbodyCoulomb} simplifies into
\begin{equation}\label{eq:Eloc11m}
  E_\mathrm{loc}^{[\varphi]}(q_3)=-\frac{5m+1}{4(m+1)}
  -\frac{m}{2(m+1)}\left(\cos\theta_1+\cos\theta_2+\frac{2}{m+1}\cos\theta_3\right)\;.
\end{equation}
The general study of appendix~\ref{sec:appendixA} applied
for~$(a_1,a_2,a_3)=(1,1,\frac{2}{m+1})$ allows to get the following
analytic bounds on~$E_0$:
\begin{subequations}
\begin{align} 0\leqslant m&\leqslant1;&\quad 
             -\frac{1}{4}-\frac{m}{8}-\frac{m(m+2)}{(m+1)^2}
             &\leqslant E_0\leqslant
             -\frac{1}{4}-\frac{m(2m+1)}{(m+1)^2}\;;
             \\
             1\leqslant m&\leqslant3;&\quad
             -\frac{1}{4}-\frac{m}{8}-\frac{m(m+2)}{(m+1)^2}
             &\leqslant E_0\leqslant
              -\frac{1}{4}-\frac{m(m+2)}{(m+1)^2}\;;
             \\
             3\leqslant m&;&\quad 
              -\frac{1}{4}-\frac{m(2m+1)}{(m+1)^2}
             &\leqslant E_0\leqslant
             -\frac{1}{4}-\frac{m(m+2)}{(m+1)^2}\;.
\end{align} 
\end{subequations}
For~$m\leqslant3$, the lower bound on $E_0$ corresponds to a
configuration where the particles make a non-degenerate isosceles
triangle whose three angles are given
by~$\cos\theta_1=\cos\theta_2=(m+1)/4$ and~$\cos\theta_3=1-(m+1)^2/8$.
The other bounds correspond to configurations where the particles are
aligned. For~$m\gg1$, the bounds saturates to~$-9/4\lesssim
E_0\lesssim-5/4$ which is quite rough compared to the numerical
value~$E_0>-1.8$ obtained for~$m=20$ with the optimized variational
method; yet it is better than the results given by the improved (Hall-Post)
variational method~\cite[Table~2]{Basdevant+93a} with which it
coincides for~$m\leqslant1$.  For small $m$ both upper and lower
bounds tend to the 2-body exact energy and provide acceptable bounds:
For instance, when~$m=0.05$, the differential method
gives~$-.34922\leqslant E_0\leqslant-.29989$ while the other
ones~\cite[Table~2]{Basdevant+93a} give~$-.59525\leqslant E_0$ (naive
variational method) , $-.34922\leqslant E_0$ (improved variational
method), $-.34666\leqslant E_0$ (optimized variational method) and
$E_0\leqslant-.3375$ (variational with hyperspherical expansion up to
$L=8$).

As already mentioned, the differential method, though being less
precise for $N=3$ than the improved or hyperspherical variational
approaches, has several advantages: it is much simpler, it provides
analytic upper and lower bounds that furnish an explicit estimation of
the errors and, at last but not least, can be easily extended to
larger~$N$ (see below) ; though possible in principle,  the generalization of the improved
variational method has not been done beyond~$N=4$.
\begin{figure}[!ht]
\begin{scriptsize}
\begin{tabular}{l|lllllllr}
\hline
$m$ & Naive & Hall-Post & Optimized & Variational & 
$E_0^{\mathrm{(d.m.)}}$ &
$\Delta E_0^{\mathrm{(d.m.)}}$&
 $E_0^{\mathrm{(d.m.)}}\mathop{-}\Delta E_0^{\mathrm{(d.m.)}}$ \\
\hline
0.05 &       -0.59525    &      -0.34922   &      -0.34666 &      -0.3375  &     -0.3246   &     0.02467 & -0.3492  \\
0.1  &       -0.6818     &      -0.436055  &      -0.43434 &      -0.423465&     -0.3926   &     0.04344 & -0.4361 \\
0.2  &       -0.8333     &      -0.58055   &      -0.58045 &      -0.55915 &     -0.5125   &     0.06806 & -0.5806 \\
0.5  &       -1.16667    &      -0.86805   &      -0.86705 &      -0.8242  &     -0.7813   &     0.08681 & -0.8681  \\
1    &       -1.5        &      -1.125     &      -1.125   &      -1.067   &     -1.0625   &     0.06250 & -1.1250  \\
2    &       -1.83333    &      -1.3889    &      -1.37135 &      -1.30225 &     -1.2639   &     0.12500 & -1.3889  \\
5    &       -2.16667    &      -1.8472    &      -1.61705 &      -1.53935 &     -1.5000   &     0.27778 & -1.7778   \\
10   &       -2.3182     &      -2.49175   &      -1.731   &      -1.6495  &     -1.6136   &     0.37190 & -1.9855   \\
20   &       -2.40475    &      -3.74775   &      -1.7972  &      -1.7134  &     -1.6786   &     0.43084 & -2.1094   \\
\hline
\end{tabular}
\end{scriptsize}
\bigskip\bigskip\bigskip

\includegraphics[width=10cm]{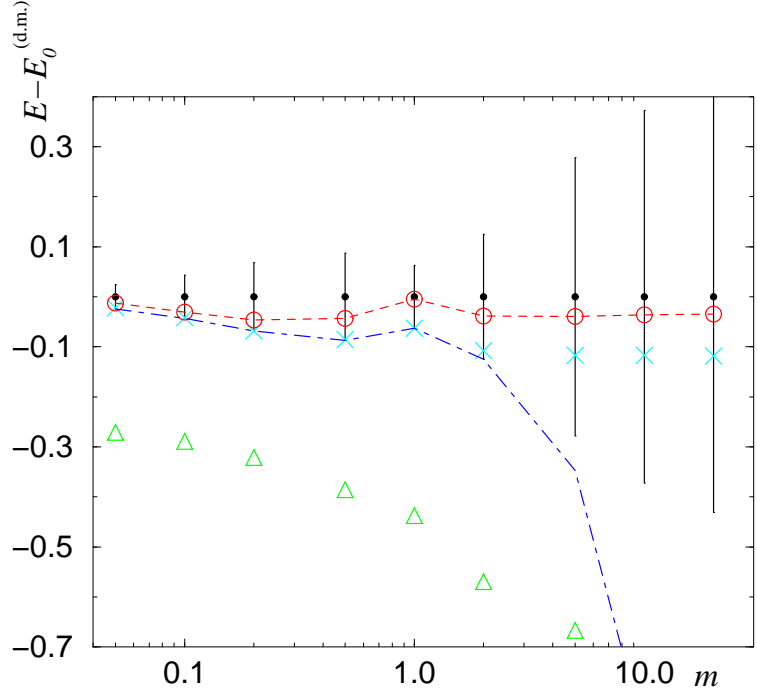}
\caption{\label{fig:Em11}Estimations for the ground-state energy~$E_0$ of an attractive Coulombian system
made of three particles with masses $(m,1,1)$. Naive ($\triangle$), Hall-Post (also
called improved, dashed-dotted link) and Optimized ($\times$) provide
lower bounds on $E_0$ while Variational
(open circles with a dashed link)
provides an upper bound. The corresponding data are taken
from\protect\cite{Basdevant+93a}. In the graph, the
value of $E_0^{\mathrm{(d.m.)}}$ has been subtracted. The error-bars
are centered at zero (filled-circles) and have a width~$2\Delta E_0^{\mathrm{(d.m.)}}$.
 }
\end{figure}

\subsection{Several examples of four-body systems}

\begin{figure}[!ht]
\begin{scriptsize}
\begin{tabular}{l|lllllllr}
\hline
$m$ & Naive & Hall-Post & Optimized & Variational & 
$E_0^{\mathrm{(d.m.)}}$ &
$\Delta E_0^{\mathrm{(d.m.)}}$ &
 $E_0^{\mathrm{(d.m.)}}\mathop{-}\Delta E_0^{\mathrm{(d.m.)}}$  \\
\hline
         0.01 &      -2.29455  &       -1.17301  &       -1.17283 &        -1.108281 &   -1.10   &  0.07   &      -1.1730 \\
          0.1 &      -2.65909  &       -1.54679  &       -1.54167 &        -1.45802  &   -1.39    &  0.16   &      -1.5468 \\
          0.5 &         -3.75  &       -2.47917  &       -2.47618 &        -2.28857  &   -2.20    &  0.28   &       -2.4792\\   
            1 &          -4.5  &             -3  &       -3       &        -2.78762  &   -2.7500 &  0.2500   &      -3.0000 \\
            3 &        -5.625  &        -3.9375  &       -3.73167 &        -3.45553  &   -3.3024 &  0.6149   &       -3.9173\\
           10 &       -6.3409  &       -6.48655  &       -4.20877 &        -3.90826  &   -3.6691 &  1.0575   &      -4.7266 \\
          100 &      -6.70545  &       -40.1395  &        -4.4673 &        -4.154310 &   -3.8412 &  1.3266   &      -5.1678 \\
          500 &        -6.741  &       -190.128  &       -4.49338 &        -4.17914  &   -3.8574 &  1.3545   &      -5.2119 \\
     $+\infty$&         -6.75  &      $ -\infty$ &       -4.5     &        -4.19259  &   -3.8615 &  1.3615   &      -5.2231\\
\hline
\end{tabular}
\end{scriptsize}
\bigskip
\includegraphics[width=10cm]{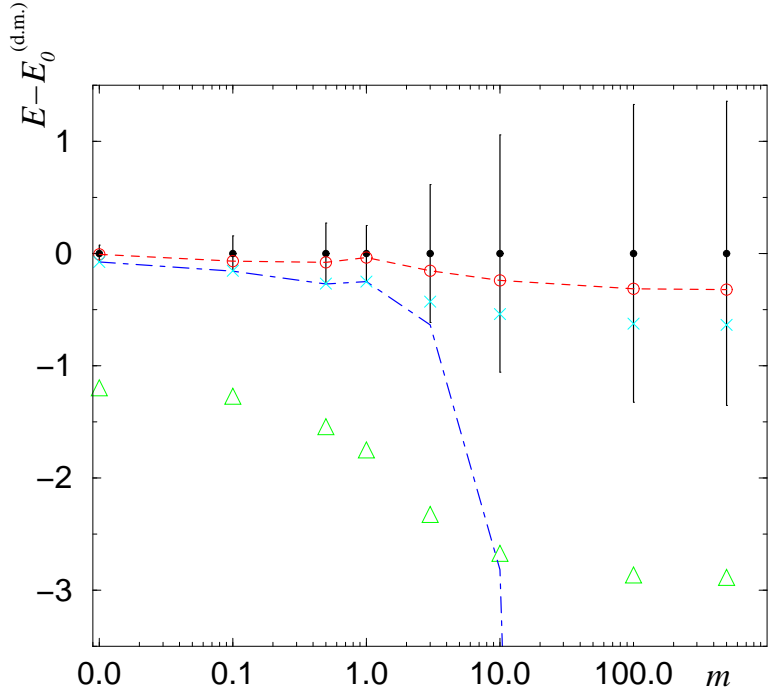}
\caption{\label{fig:Em111} Same conventions as in
  figure~\ref{fig:Em11} for an attractive Coulombian system
made of four particles with masses $(m,1,1,1)$.
}
\end{figure}

\begin{figure}[!ht]
\begin{scriptsize}
\begin{tabular}{l|lllllllr}
\hline
$m$ & Naive & Hall-Post & Optimized & Variational & 
$E_0^{\mathrm{(d.m.)}}$ &
$\Delta E_0^{\mathrm{(d.m.)}}$ &
 $E_0^{\mathrm{(d.m.)}}\mathop{-}\Delta E_0^{\mathrm{(d.m.)}}$  \\
\hline
           0.001 &       -0.756745 &       -0.504495 &        -0.25557 &        -0.25492 &  -0.2543   &  0.0021  &    -0.2564   \\
           0.002 &       -0.763475 &       -0.508985 &        -0.26114 &        -0.25985 &  -0.2587   &  0.0042  &    -0.2629  \\
           0.005 &         -0.7836 &         -0.5224 &       -0.277805 &         -0.2746 &  -0.2717   &  0.0104  &    -0.2820    \\
            0.01 &       -0.816905 &       -0.544605 &       -0.305465 &        -0.32403 &  -0.2931   &  0.0204  &    -0.3135    \\
            0.05 &        -1.07322 &       -0.715475 &       -0.519635 &        -0.50503 &  -0.4581   &  0.0913  &    -0.5494     \\
             0.1 &        -1.37045 &       -0.913635 &        -0.76439 &         -0.7308 &  -0.6492   &  0.1594  &    -0.8086   \\
             0.2 &            -1.9 &        -1.26666 &        -1.17921 &        -1.10975 &  -0.9893   &  0.2448  &    -1.234   \\
             0.5 &          -3.125 &        -2.08333 &        -2.06426 &        -1.91867 &  -1.7847   &  0.2986  &    -2.0833$^{\dagger}$ \\
             0.8 &        -4.01666 &        -2.67778 &        -2.67552 &        -2.48094 &  -2.4034   &  0.2744  &    -2.6778$^{\dagger}$ \\
               1 &            -4.5 &              -3 &              -3 &        -2.78736 &  -2.7500   &  0.2500  &    -3.0000$^{\dagger}$       \\
\hline
\end{tabular}
\end{scriptsize}
\bigskip
\includegraphics[width=10cm]{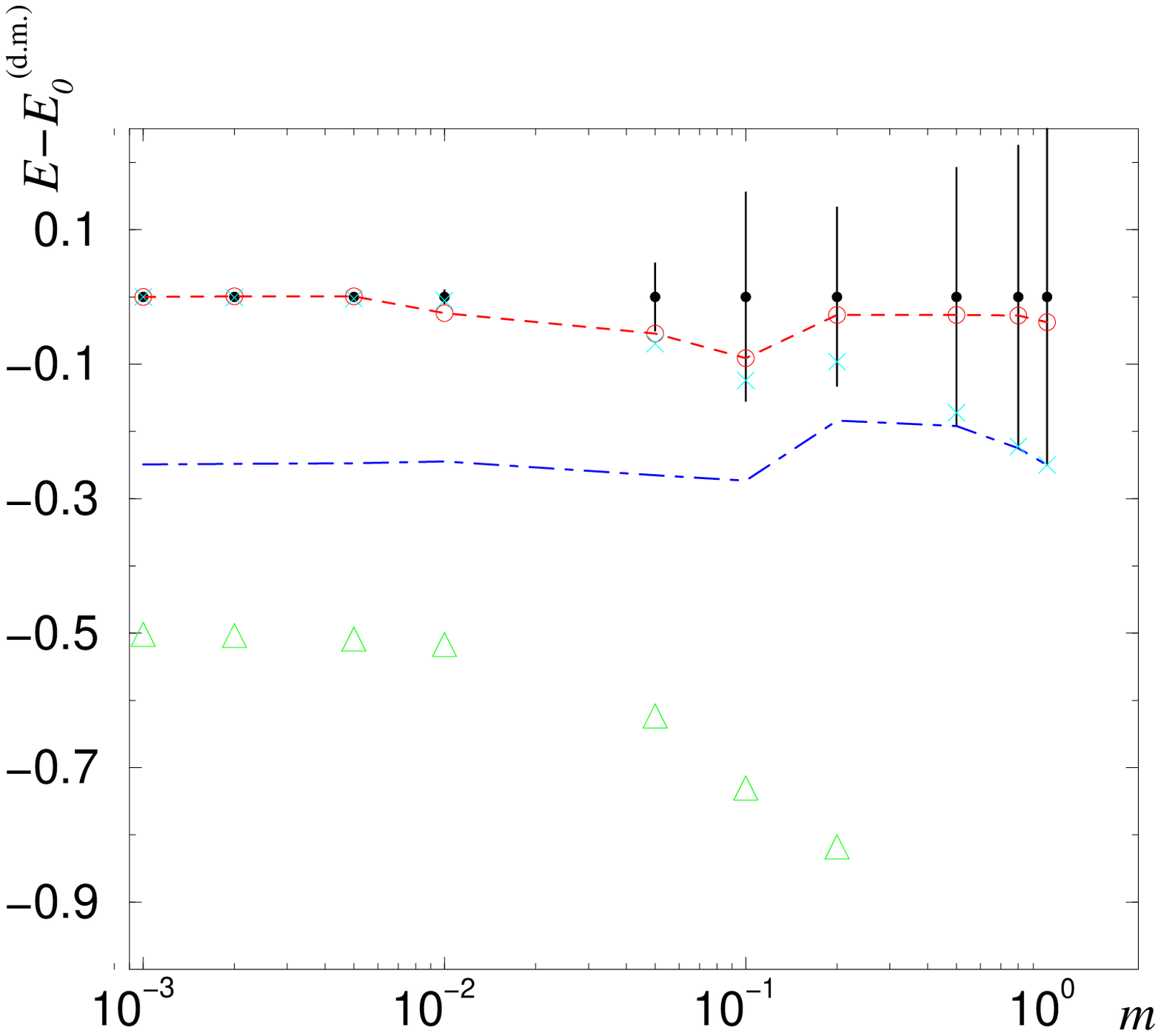}
\caption{\label{fig:Emm11}Same conventions as in
  figure~\ref{fig:Em11} for an attractive Coulombian system
made of four particles with masses $(m,m,1,1)$. Numbers signaled with a $^{\dagger}$ are the lower 
bounds~\eqref{eq:Emaxmm11}.}
\end{figure}

The optimized variational method has been successfully proposed
for~$N=4$ in \cite{Benslama+98a} for potentials with scaling-law
behavior. For Coulombian interactions with a common coupling constant
set to $-1$, tables and figures~\ref{fig:Em111} and \ref{fig:Emm11} compare the variational results to
those obtained from the differential method when~$\textsc{d}=3$.  The same
conclusion as in the previous section can be drawn and here are some
examples of explicit analytic bounds that are obtained by partitioning
$F_4$ in subclusters made of 3 particles:

Let us take $m_1=m_2=m_3=1$ and $m_0=m$. We have
\begin{subequations}
\begin{equation}\label{eq:Em111}
 0 \leqslant  m\leqslant 1; \quad
 -\frac{9}{8}-\frac{3m(m^2+6m+13)}{8(m+1)^2}
  \leqslant E_0\;;
\end{equation}
\begin{multline}\label{eq:Em111_bis}
  1\leqslant m\leqslant2\sqrt{3}-1;\\
 -\frac{9}{8}-\frac{3m(m^2+6m+13)}{8(m+1)^2}
  \leqslant E_0\leqslant
  -\frac{5m^2+13m+2}{2(m+1)^2}\;;
\end{multline}
\begin{multline}\label{eq:Em111_ter}
  2\sqrt{3}-1\leqslant m; \\  
 -\frac{9}{8}-\frac{3m(m^2+6m+13)}{8(m+1)^2}+\frac{3m(m+1-2\sqrt{3})^2}{8(m+1)^2}
  \leqslant E_0
   \leqslant
  -\frac{5m^2+13m+1}{2(m+1)^2}\;.
\end{multline}
\end{subequations}

The configuration that minimizes
$E_\mathrm{loc}^{[\varphi]}(q_4)$ given by~\eqref{eq:ElocNbodyCoulomb}
corresponds to a tetrahedron with an equilateral basis made by
particles 1, 2 and 3. The three other faces, with particle $0$ at one
vertex, are identical isosceles triangles, namely those which maximize
\eqref{eq:Eloc11m} when $m\leqslant3$. Such a tetrahedron can indeed
be constructed provided the angles at particle $0$ are lower than~$2\pi/3$
which requires~$m\leqslant2\sqrt{3}-1\simeq2.464$. For $m\geqslant2\sqrt{3}-1$,
the configuration that minimizes $E_\mathrm{loc}^{[\varphi]}(q_4)$ seems numerically
to correspond to an equilateral triangle made  by
particles 1, 2 and 3 with particle $0$ at its center (the flat tetrahedron obtained 
when~$m=2\sqrt{3}-1$). This simple configuration allows to conjecture the analytic
lower bound in \eqref{eq:Em111_ter}. For $m\geqslant1$, 
the upper bound is obtained when the four particles
are aligned with the~\nth{0} at one extremity.  For $m\leqslant1$, I am not able to 
propose an analytic expression for the upper bound. 

\begin{figure}[!ht]
\center
\includegraphics[width=5cm]{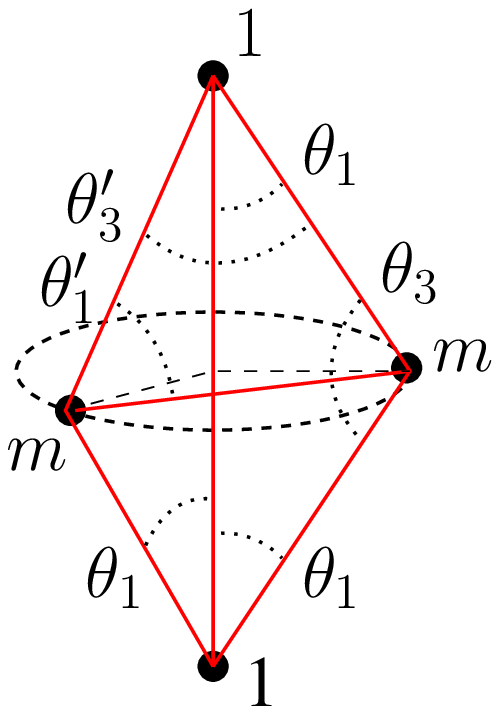}
\caption{\label{fig:config_mm11} For a special range of $m$, a
  four-body configuration with masses $(m,m,1,1)$ that minimizes the
  local energy can be constructed from the optimal configurations
  of each of its three-body subsystems. }
\end{figure}

For $m_2=m_3=1$ and $m_0=m_1=m$, a tetrahedron that maximizes all the
contributions of its faces simultaneously can be constructed for
$m_\star^{-1}\leqslant m\leqslant m_\star\DEF
(-1+\sqrt{17}+\sqrt{14-2\sqrt{17}})/2$.  Two identical faces (see
figure
~\ref{fig:config_mm11})
corresponding to particles with masses $(1,1,m)$ have their three
angles given by~$\cos\theta_1=\cos\theta_2=(m+1)/4$
and~$\cos\theta_3=1-(m+1)^2/8$. The angles $(\theta_1'=\theta_2',
\theta_3')$ of the two other faces corresponding to particles with
masses $(m,m,1)$ are obtained replacing $m$ by $m^{-1}$ in the
previous expressions. For $1/3\leqslant m\leqslant3$ such faces can
indeed be constructed but the pairs of identical faces can be put
together to construct one tetrahedron provided only that
$\theta_3\leqslant2\theta_1'$. This last condition leads to
$m^4+2m^3-14m^2+2m+1\leqslant0$. $m_\star$ and~$m_\star^{-1}$ are the
two positive roots of the four-degree-polynomial, the two others being
negative.  We get
\begin{equation}\label{eq:Emaxmm11}
  m_\star^{-1}\simeq.3622\leqslant m\leqslant m_\star\simeq2.7609;\quad
 -\frac{m^2+10m+1}{2(m+1)}\leqslant E_0\;.
\end{equation}

\subsection{Arbitrary number of identical particles plus one different from the others}

As we have already seen for identical particles, the possibility of
partitioning the local energy in contributions involving more than two
particles allow to get some bounds on the ground-state energy of
systems made of an arbitrary number~$N$ of
particles. For~$N\geqslant5$, this is beyond the scope of the existing
optimized variational methods. To see one more last example, let us
generalize both cases of sections~\ref{subsec:Nidentical} and
\ref{subsec:11m} and consider a system made of one particle with
mass~$m_0=m$ and~$N-1$ identical particles of
mass~$m_1=\cdots=m_{N-1}=1$.  All the identical particles interact
with the same coupling constants~$e_{ij}=-1$ and the \nth{0} interact
with $e_{0,i}=-g^2$. The local energy is given
by~\eqref{eq:ElocNbodyCoulomb} as
\begin{multline}
   E_\mathrm{loc}^{[\varphi]}(q_N)=-\frac{1}{(\textsc{d}-1)^2} \Bigg\{
\frac{2mg^4}{m+1}(N-1)+\frac{1}{2}(N-1)(N-2)+F_{N-1}(q_{N-1})\\
  +a(m^{-1})\sum^{N-1}_{\substack{i,j=1\\i<j}}\left[\cos(\widehat{0ij})+\cos(\widehat{0ji})+a(m)\cos(\widehat{i0j})
                                      \right]
  \Bigg\}
\end{multline}
where now~$q_{N-1}$ stands for the configuration of the $N-1$
identical particles. $a(m)\DEF2g^2/(m+1)$.  By simultaneously bounding
the contributions of the triangles that include the \nth{0} particle
and the contribution of the remaining cluster of the identical
particles, we immediately have analytic expressions for bounds
on~$E_0$. For instance the lower bound is given by
\begin{multline}
  E_0\geqslant-\frac{1}{(\textsc{d}-1)^2} \Bigg\{
\frac{2mg^4}{m+1}(N-1)+\frac{1}{2}(N-1)(N-2)+\sup_{\mathcal{Q}_{N-1}}F_{N-1}\\
 +\frac{1}{2}\,a(m^{-1})(N-1)(N-2)
F_3^\mathrm{max}\big(1,1,a(m)\big) \Bigg\}
\end{multline}
where~$F_3^\mathrm{max}$ is given  by~\eqref{eq:F3minmax} and 
$\sup_{\mathcal{Q}_{N-1}}F_{N-1}$ has been explicitly estimated in section~\ref{subsec:Nidentical}.

\section{Arbitrary two-body  interaction}\label{sec:generalvij}

\subsection{Behavior at large distances}

Considering attractive  Coulombian interactions is relevant for heavy quarks
models at short distances but, of course, other
kinds of effective potentials are required in most models.  Since in
general no analytic expressions are known for the two-body 
ground-state energies~$\epsilon_{ij}$, no method is expected to provide
explicit non-trivial bounds on~$E_0$.  However, if one has some
experimental clues about $\epsilon_{ij}$ (by measuring 2-body masses
or dissociation energies) or numerical estimates as well, it is always
interesting to obtain some relations between the~$\epsilon$'s and the
ground-state energies of larger systems.  As mentioned in the
introduction, this have been achieved in \cite{Basdevant+90b} for
$N=3$ and in \cite{Benslama+98a} for $N=4$ when the interactions are
of the form $v_{ij}(r)\propto\sign(\beta)r^\beta$.

For~$\beta>0$, the semiclassical argument given at the end of
section~\ref{sec:diffmethod} shows that \eqref{eq:NbodyElocSSradial}
is expected to be unbounded; then \eqref{eq:inequalities} gives no
information. If we want to take the advantage of the simple form
\eqref{eq:NbodyElocSSradial} (that is, to keep the
choice~\eqref{eq:totalphi} with~\eqref{eq:schro2body} for the test
functions), we have to work with finite range potentials.  When at
large distances, the potential is still confining ($\beta=2$ for
harmonic forces or $\beta=1$ for interquark force in quantum
chromodynamics \cite{Grant+93a} and \cite[for an up-to-date
review]{Brambilla+04a}), some different ansatz for~$\varphi$ must be
constructed in order to eliminate the singular behavior of the $v_{ij}$'s at
infinite distances. Actually, the Coulombian case considered in the
previous section can be seen as an example of a problem where simple
poles at finite distances can be eliminated. Anyway, in many
situations, the 2-body ground-state is expected to depend on the
behavior of the potential at large distances by exponentially small
terms only. If, in the integral \eqref{eq:integralidentity}, we decide
to keep only those configurations~$q_N$ whose size remains in a
physical domain bounded by a cut-off length~$\Lambda$, then we expect to
make an exponentially small error on the estimates of $E_0$; this is due to
the exponential decay of~$\Phi_0$ when two or more particles separate
off.  Like the $\epsilon_{ij}$'s, $\Lambda$ is typically obtained from
a 2-body dynamics but its precise value is irrelevant if the extremal
values of the local energy do not depend on it. It is precisely the
case of the Coulombian interactions (more exactly,  interactions that
can be modeled by Coulombian potential at the energy scale where the
ground-state exists) for which the local energy
\eqref{eq:ElocNbodyCoulomb} is invariant under dilations.

 Since, in the present section, we just want to sketch some main
 guidelines without working through the details neither being
 exhaustive, we will consider only the cases where
\begin{equation}\label{eq:vinfinity}
  v(r)\underset{r\to+\infty}{\longrightarrow}0\;.
\end{equation}

\subsection{Fitting the 2-body ground-state wavefunction}

What is new, here, is that the differential method allows us to choose
directly the 2-body ground-state wave-functions, or rather their
logarithms~$S_{ij}$.  Once some numerical estimate of~$\epsilon_{ij}$
is obtained in one way or another, we can completely bypass the
problem of modeling the 2-body potential uniformely. Being free of any
integration, the differential method can deal with rather complicated,
and therefore rather realistic two-body test functions. An explicit
choice of~$S_{ij}$'s provides an explicit form for the local energy
\eqref{eq:NbodyElocSSradial}.

It frequently happens that we know from experiments the behavior of
the two-body potential in some specific regimes (most generally at
short and large distances) but not uniformly.  We can therefore, in
each of these regimes, tentatively obtain, with the help of the
differential equation~\eqref{eq:schro2body}, the local functional form
of the two-body ground-state wave-function. Matching these local
solutions together, and then dealing with quite complicated global
expression for~$S$ and $v$, do not represent a serious obstacle for
the computation of \eqref{eq:NbodyElocSSradial}.

To be a little less speculative, let us consider $N$ identical particles with unit mass,
 interacting with a two-body radial potential~$v(r)$ such that
$v(r)\to0$ when $r\to\infty$ and
\begin{equation}\label{eq:v0plus}
  v(r)\underset{r\to0^+}{\longrightarrow}v_0\,r^\beta
\end{equation}
for some $v_0$ and with considering only one case, say $\beta>0$.     
 The two-body stationary Schr\"odinger
equation~\eqref{eq:schro2body} becomes
\begin{equation}\label{eq:schro2bodyradial}
  -\left(\frac{\ud^2}{\ud r^2}+\frac{\textsc{d}-1}{r}\frac{\ud}{\ud r}\right)\phi+v\phi=\epsilon_0\phi
\end{equation}
where
 $\epsilon_0<0$ will denote an estimate
of the two-body ground-state energy; it can be considered as another
parameter that should fit the experiments involving two bodies.
Let us guess the behavior of~$S(r)\DEF\ln\phi(r)$ at short
distances by writing for $\sigma\neq-1$:
\begin{equation}\label{eq:S0plus}
  S(r)\underset{r\to0^+}{\longrightarrow}\frac{s_0}{\sigma+1}\,r^{\sigma+1}\;.
\end{equation}
 Identifying
the leading orders after having reported \eqref{eq:S0plus} 
in \eqref{eq:schro2bodyradial}, we necessarily get (for $\textsc{d}>1$)
$\sigma=1$ and $s_0=-\epsilon_0/\textsc{d}$. The next term in the development 
of~$S$ can also be determined. For $0<\beta<2$, it depends only on
the leading term \eqref{eq:v0plus} and we have
\begin{equation}
  S(r)\underset{r\to0^+}{\longrightarrow}-\frac{\epsilon_0}{2d}\,r^2
  +\frac{v_0}{(\textsc{d}+\beta)(2+\beta)}\,r^{2+\beta}+o(r^{2+\beta})\;.
\end{equation} 
This local asymptotic series must be matched with the semiclassical
behavior at large~$r$ 
\begin{equation}
  S(r)\underset{r\to+\infty}{\longrightarrow}-\sqrt{-\epsilon_0}\,r
\end{equation} 
since we have supposed \eqref{eq:vinfinity}. The additive constant
in $S$
is irrelevant since the local energy does not depend on
the normalization of $\phi$.
A simple choice that ensures the local energy to remain uniformly bounded is
to take  for $S'$ a  fraction like
\begin{equation}\label{eq:ansatzSp}
  S'(r)=\frac{-\frac{\epsilon_0}{\textsc{d}}\,r+\frac{v_0}{\textsc{d}+\beta}\,r^{1+\beta}
  -\sqrt{-\epsilon_0}\,r^{1+2\beta}}{1+r^{1+2\beta}}\;.
\end{equation}
\begin{figure}[!ht]
\center
\includegraphics[width=12cm]{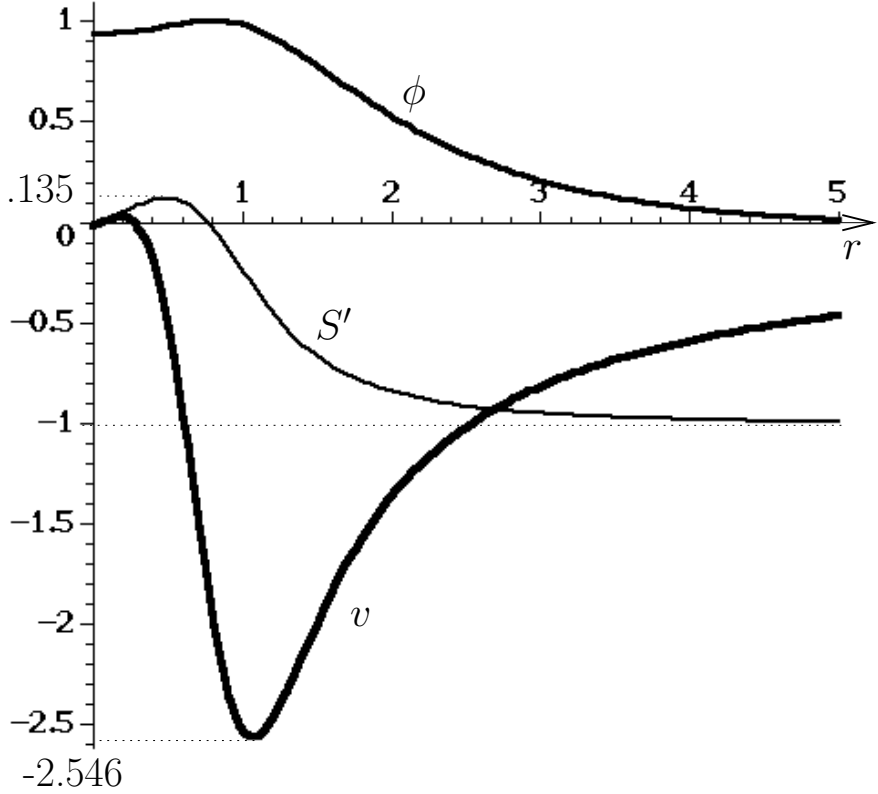}
\caption{\label{fig:vphiSp}  $\textsc{d}=3$, $\epsilon_0=-1$, $v_0=1$, $\beta=1.5$, the 
function~$S'$ given by~\eqref{eq:ansatzSp} is plotted as well as the corresponding wavefunction $\phi=\exp{S}$ which is
the ground-state of the potential~$v$ for energy~$\epsilon_0$.
 The latter can be reconstructed from~$\phi$ with the Schr\"odinger equation.  
}
\end{figure}

Figure~\ref{fig:vphiSp} show the corresponding $\phi$ for an arbitrary choice
of parameters together with the corresponding
$v$ whose  complicated analytic expression on $[0,+\infty[$ is not
needed.

\subsection{Crude bounds}

 From equation~\eqref{eq:NbodyElocSSradial}, an immediate upper bound on $E_0$ is given by:
\begin{multline}\label{eq:roughwindow}
  \frac{1}{2}N(N-1)\,\epsilon_0-\frac{1}{2}N(N-1)(N-2)\,\sigma^2
  \leqslant E_0 \\
  \leqslant
  \frac{1}{2}N(N-1)\,\epsilon_0+\frac{1}{2}N(N-1)(N-2)\,\sigma^2
\end{multline}
where 
\begin{equation}\label{def:sigma}
  \sigma\DEF\sup_{[0,+\infty[}|S'|\;.
\end{equation}

 Because the constraints between
the angles~$\widehat{j,i,k}$ are not taken into account, these inequalities are expected to be rather rough
and their quality deteriorate for large~$N$:
when positive, the upper bound becomes irrelevant since we already know that~$E_0\leqslant0$.
Indeed, the decreasing of the $\sigma^2$ term with $N$ does not guarantee
that the lower bound is better than the Hall-Post bound $N(N-1)\tilde{\epsilon}_0/2$
or even than the naive one $N(N-1)\tilde{\tilde{\epsilon}}_0/2$ \cite[\S~2.1 and 2.2]{Benslama+98a}
 where  $\tilde{\tilde{\epsilon}}_0$
(resp. $\tilde{\epsilon}_0$ and $\epsilon_0$)
is the ground-state energy for a particle of mass~$(N-1)/2$ (resp. $N/4$ and 1/2) in the central potential $v$
(recall $\tilde{\tilde{\epsilon}}_0<\tilde{\epsilon}_0<\epsilon_0$).

\subsection{Reduction to a finite number of Coulombian cases}

In fact, we can find some bounds of~\eqref{eq:NbodyElocSSradial} by 
reducing the problem to  
a finite number of Coulombian-like cases, that is,
  where the function to be bound involves constant factors 
in front of the cosine
(compare \eqref{eq:ElocNbodyCoulomb} to \eqref{eq:NbodyElocSSradial}).
 To see this, split the 
coordinates~$q_N$ into a scaling factor~$\lambda\geqslant0$ and some
 angle variables~$\theta$ among which $(N-1)\textsc{d}-1$ are independent. Each distance 
writes~$r_{ij}=\lambda\rho_{ij}(\theta)$ 
where the $\rho_{ij}$'s are functions that
do not depend on the global size of the configuration but on its shape
only.  Now, from~\eqref{eq:NbodyElocSSradial},
  we define (recall $m_i=1$)
\begin{equation}\label{def:GN}
   G_N(q_N)\DEF\sum_{(\widehat{j,i,k})}S'(r_{ij})S'(r_{ik})\cos(\widehat{j,i,k})
\end{equation}
and we have
\begin{equation}
  \inf_{q_N}\big(G_N(q_N)\big)=\inf_{\theta}\big(\tilde{G}_N(\theta)\big)
\end{equation}
where
\begin{equation}
\tilde{G}_N(\theta)=\inf_{\lambda}\sum_{(\widehat{j,i,k})}S'[\lambda \rho_{ij}(\theta)]\,S'[\lambda \rho_{ik}(\theta)]\cos(\widehat{j,i,k})\;.
\end{equation}
Analogous relations are obtained for the maxima.
For fixed~$\theta$, when $\lambda$ varies from $0$ to $+\infty$, the map
 $\lambda\mapsto\big(S'[\lambda \rho_{ij}(\theta)]\big)_{0\leqslant i<j\leqslant N-1}$
 defines a curve~$\mathcal{C}_\theta$ in a $n$-dimensional space with $n=N(N-1)/2$. 
$\mathcal{C}_\theta$ is bounded if $S'$ is bounded. More precisely, $\mathcal{C}_\theta$ is 
inside the $n$-dimensional hypercube 
$\mathcal{B}\DEF[\sigma_{\mathrm{min}},\sigma_{\mathrm{max}}]^n$ where 
\begin{subequations}\label{def:sigmaminmax}
\begin{equation}
  \sigma_{\mathrm{min}}\DEF\inf_{[0,+\infty[}S'
\end{equation}
and
\begin{equation}
  \sigma_{\mathrm{max}}\DEF\sup_{[0,+\infty[}S'\,.
\end{equation}
\end{subequations}
If $S'$ has the form shown in figure \ref{fig:vphiSp}, $\mathcal{C}_\theta$
 starts at the origin ($S'(0)=0$) and ends at the point
 $(-1,\dots,-1)$.  Taking all the points in $\mathcal{B}$ rather than
 the points of~$\mathcal{C}_\theta$ leads to a lower bound of $\tilde{G}_N$:
\begin{equation}\label{eq:BGN}
  \inf_{\vec{s}\in\mathcal{B}}\sum_{(\widehat{j,i,k})}s_{ij}\,s_{ik}\cos(\widehat{j,i,k})
  \leqslant \tilde{G}_N(\theta)
\end{equation}   
where $\vec{s}=(s_{ij})_{0\leqslant i<j\leqslant N-1}$. Now, whatever
the values of the cosines may be, the quadratic function in~$\vec{s}$
appearing in the left-hand side of~\eqref{eq:BGN} reaches its minimum
at a vertex of~$\mathcal{B}$\footnote{For any constant $A$, any
$n$-dimensional vector $\vec{B}$ and any symmetric $n\times n$-matrix
$C$ with vanishing diagonal coefficients, the critical points of
$f_n(\vec{s}\,)=\vec{s}\cdot C\vec{s}+\vec{B}\cdot\vec{s}+A$ are
always saddle points: the direction $s_2=\pm\sign(C_{12})s_1$ and
$s_i=0$ for $i>2$ makes $f$ increase/decrease as $\pm|C_{12}|s_1^2$
from its critical value. Therefore the extrema of $f$ are reached on
the boundary of the domain of $\vec{s}$. For $\vec{s}$ restricted to a
$n$-dimensional squared box whose faces are given by fixing one $s_i$,
 the restriction of $f_n$ to one face,
\textit{i.e,} to a $(n-1)$-dimensional box, leads to a
function~$f_{n-1}$ to which the above argument may be applied
again. By repetition down to $n=1$, we see that the maximum and the
minimum of $f$ is necessarily reached at one of the vertices of the
original $n$-box.  }. Let us denote by $\mathcal{V}$, the finite set
of the $2^n$ vertices $\vec{s}$ of $\mathcal{B}$ (\textit{i.e.} for all $\vec{s}$
in $\mathcal{V}$, each $s_{ij}$ is either $\sigma_{\mathrm{min}}$ or
$\sigma_{\mathrm{max}}$). We have
\begin{equation}\label{eq:infvertex}
    \inf_{\vec{s}\in\mathcal{V}}\big[\inf_\theta\big( F^{(\vec{s}\,)}_N(\theta)\big)\big]
\leqslant\inf_{q_N}\big(G_N(q_N)\big)\;.
\end{equation} 
where
\begin{equation}\label{def:FNs}
   F^{(\vec{s}\,)}_N(\theta)\DEF\sum_{(\widehat{j,i,k})}s_{ij}s_{ik}\cos(\widehat{j,i,k})\;.
\end{equation}
In fact, what we have done  by obtaining the left-hand side
 of~\eqref{eq:infvertex} is to 
make the values of~$S'(r_{ij})$ independent from those of
$\theta$. It follows that the inequality~\eqref{eq:infvertex} will be
strict if the value of~$\vec{s}$ at the minimizing vertex are
incompatible with the geometrical constraints on the configuration of
the $N$ points. We will illustrate this point in the next subsection.
We have obtained
\begin{subequations}\label{subeq:finitevertices}
\begin{equation}
  E_0\leqslant\frac{1}{2}N(N-1)\epsilon_0-\inf_{\vec{s}\in\mathcal{V}}\big[\inf_\theta\big( F^{(\vec{s}\,)}_N(\theta)\big)\big]\;.
\end{equation} 
and similarly
\begin{equation}
 \frac{1}{2}N(N-1)\epsilon_0-\sup_{\vec{s}\in\mathcal{V}}\big[\sup_\theta\big( F^{(\vec{s}\,)}_N(\theta)\big)\big]\leqslant E_0\;.
\end{equation}
 \end{subequations}
The function~$F^{(\vec{s}\,)}_N(\theta)$
has the same form as the second sum of the right-hand of~$\eqref{eq:ElocNbodyCoulomb}$. Therefore,
the computation of the bounds in~\eqref{subeq:finitevertices}  is equivalent to a finite number of Coulombian
problems (with not necessarily attractive interactions since the sign of~$s_{ij}$ may change)
 where we must consider all the possible~$\vec{s}$
whose components are either
 $\sigma_{\mathrm{min}}$ or $\sigma_{\mathrm{max}}$.

\subsection{Three bodies}

As we have seen in section~\ref{sec:coulomb}, even in the purely attractive  Coulombian cases, an
analytic expression of the extrema of~$F_N$ is not known in general. Anyway, one can always
group in clusters the terms involved in \eqref{def:FNs} like in~\eqref{eq:clustering},  then use inequalities like~\eqref{eq:alphaMalphaN}
and reduce the number of particles.
Let us then consider $N=3$. It can be 
shown\footnote{\label{fn:exG3} The extrema of
$F^{(\vec{s}\,)}_3(\theta)$ when $\theta$ varies can be calculated with the help of the appendix with
$a_3\mathop{=}s_{01}s_{02}$, $a_2\mathop{=}s_{02}s_{12}$ and~$a_1\mathop{=}s_{01}s_{12}$. From \eqref{def:f}, we get
$f(a_1,a_2,a_3)=\frac{1}{2}(s_{01}^2+s_{02}^2+s_{12}^2)$ which is  always positive and larger than
$a_1+a_2-a_3$, $-a_1+a_2+a_3$ and  $a_1-a_2+a_3$ corresponding to the
three aligned configurations for different ordering of the particles. From \eqref{eq:F3minmax}, the minimum of~$F^{(\vec{s}\,)}_3(\theta)$
must therefore correspond to an aligned configuration. Its maximum is reached for
the configuration described just after equation~\eqref{eq:supG3}.} that
\begin{equation}\label{eq:2sigmaminmax-sigma}
  \inf_{\vec{s}\in\mathcal{V}}\big[\inf_\theta\big( F^{(\vec{s}\,)}_3(\theta)\big)\big]=2\sigma_{\mathrm{min}}\sigma_{\mathrm{max}}-\sigma^2
\end{equation}
with definitions~\eqref{def:sigma} and~\eqref{def:sigmaminmax}.
$\inf_\theta F^{(\vec{s}\,)}_3(\theta)$ is obtained for an aligned configuration which is generically incompatible
with~$\vec{s}$ being a vertex of the cube~$[\sigma_{\mathrm{min}},\sigma_{\mathrm{max}}]^3$. For instance,
suppose that~$S'$ has the shape depicted in figure~\ref{fig:vphiSp} 
where~$\sigma=-\sigma_{\mathrm{min}}>\sigma_{\mathrm{max}}>0$; the 
value
$2\sigma_{\mathrm{min}}\sigma_{\mathrm{max}}-\sigma_{\mathrm{min}}^2$
is obtained for~$\vec{s}=(s_{01},s_{02},s_{12})=(\sigma_{\mathrm{min}},\sigma_{\mathrm{min}},\sigma_{\mathrm{max}})$ and should be
 realized for~$r_{01}\gg1$, $r_{02}\gg1$ and~$r_{12}\simeq r_{\mathrm{max}}$ (the unique finite  distance at which $S''$
vanishes); but this is incompatible with the alignment
condition $\cos(\widehat{1,0,2})=-1$ where particle $0$ is in between the two others which implies
 $r_{12}=r_{01}+r_{02}$. The inequality~\eqref{eq:infvertex} is therefore strongly strict. It 
can be improved by reducing the size of the cube~$\mathcal{B}$ to make its minimizing vertices
compatible with the aligned configuration. It can be shown that  for~$S'$
of the form shown in figure~\ref{fig:vphiSp}, we have
\begin{equation}\label{eq:infG3compat}
   2\sigma_{\mathrm{max}}S'(2r_0)-\sigma_{\mathrm{max}}^2\leqslant\inf_{q_3}\big(G_3(q_3)\big)
\end{equation}  
where~$r_0$ is the unique strictly positive distance where~$S'$ 
vanishes\footnote{Any aligned configuration
with~$r_{01}\simeq r_{02}\lesssim r_0$ and~$r_{12}=r_{01}+r_{02}>r_0$ corresponds 
to a negative~$G_3$. Therefore, as far as its minimum is concerned, the configurations leading
to a positive~$\inf_\theta F^{(\vec{s}\,)}_3$ can be forgotten (see note~\ref{fn:exG3}).
It is straightforward to check that all the possible 
 relative positions of~$r_{ij}$ with respect to $r_\mathrm{max}$ and~$r_0$ that are compatible with~$r_{12}=r_{01}+r_{02}$
provide a~$G_3$ such that~\eqref{eq:infG3compat}. 
}. Even though
the inequality is still strict because $r_{12}=2r_0\neq r_{01}+r_{02}=2r_{\mathrm{max}}$ in general,
the bound is much better than~\eqref{eq:2sigmaminmax-sigma}. For instance, if we take the value of the parameters
corresponding to figure~\ref{fig:vphiSp} we have 
\begin{equation}
  2\sigma_{\mathrm{min}}\sigma_{\mathrm{max}}-\sigma_{\mathrm{min}}^2\simeq-1.2705<
  2\sigma_{\mathrm{max}}S'(2r_0)-\sigma_{\mathrm{max}}^2\simeq-.205\;.
\end{equation}
to be compared with the result of the numerical minimization of~$G_3$ 
\begin{equation}\label{eq:infG3}
  \inf_{q_3}\big(G_3(q_3)\big)\simeq-.1150\;.
\end{equation}
obtained for the aligned configuration where $r_{01}=r_{02}=r_{12}/2\simeq.6107$.

The other bound
\begin{equation}\label{eq:supG3}
  \sup_{q_3}\big(G_3(q_3)\big)=\sup_{\vec{s}\in\mathcal{V}}\big[\sup_\theta\big( F^{(\vec{s}\,)}_3(\theta)\big)\big]
=\frac{3}{2}\,\sigma^2
\end{equation}
 is actually obtained at the vertex~$\vec{s}=(\sigma,\sigma,\sigma)$ for an equilateral configuration
where the common distance~$r_{01}=r_{02}=r_{12}$ is where~$|S'|$ reaches its maximum.
For~$S'$ like in figure~\ref{fig:vphiSp}, it corresponds to a very large triangle
($r_{01}\gg1$) where~$\sigma\simeq1$. 
For $\textsc{d}=3$, $\epsilon_0=-1$, $v_0=1$, $\beta=1.5$, from~\eqref{eq:infvertex}
and results
\eqref{eq:infG3}, \eqref{eq:supG3}, the inequalities
\eqref{subeq:finitevertices} give
\begin{equation}
-3-3/2=-4.5 \leqslant E_0\leqslant -3+.1150=-2.885
\end{equation}
corresponding to a relative error of~$\Delta E_0^{\mathrm{(d.m.)}}/E_0^{\mathrm{(d.m.)}}\simeq22\%$.
This not really impressive (again we emphasize that we are not looking
for numerical performance at this stage of development of the
differential method) but can be seen as an encouraging starting point
since the interactions involved so far in the three-body system are
highly not trivial.  It would have required much more numerical work
to obtain a rigorous window for $E_0$ with variational methods
(specially a lower bound since the potential considered here does not
follow a power-law behavior).

\section{Conclusion} 
 
The differential method appears to offer a completely new strategy for
estimating a ground-state energy. For many-body systems, we have seen
on several examples how this approach can be fruitful.  For attractive
Coulombian particles, it can compete with existing others methods
(that are based on the variational principle) on several levels: it
provides upper \emph{and} lower bounds with comparable numerical
precision, its simplicity renders the analytic calculations tractable
even for large $N$ and/or allows a low cost of numerical computation.
Beyond purely Coulombian systems, the differential method, being so
general, offers a remarkable flexibility. As have been sketched in the
previous section, one can deal with systems where interactions can be
very rich (possibly short-ranged with an a priori cut-off);
 several regimes which are valid at different scales can be
implemented at once. There is some hope that future works successfully
apply the differential methods for proper realistic potentials.

Unfortunately, I have not been able to generalize the differential
method to fermionic systems where the ground-state spatial
wave-function is antisymmetric. In such cases, the presence of
non-trivial nodal lines \cite{Ceperley91a} breaks down the proof of
inequalities~\eqref{eq:inequalities}.

There is a lot of work to be done regarding a systematic improvement of
the bounds, once some finite ones have been found with a $\varphi$
given at first attempt. In this paper, we have not considered some
free parameters on which a (family) of test functions, say
$\varphi_\zeta$, may depend.  As shown for a one-dimensional system
\cite{Mouchet05a}, the locality of the differential method may require
a very few number of $\lambda$'s at each optimization step (unlike for
the variational test functions) for obtaining substantial improvements
of the bounds by calculating, say $\sup_{\zeta}\big[ \inf_{q}
\big(E_\mathrm{loc}^{[\varphi_\zeta]}(q)\big)\big]$. A precise proof that
this approach is efficient for several dimensions remains an open
interesting problem.
 
I thank Jean-Marc Richard for a critical reading of the first proof of this manuscript
and acknowledge
the generous hospitality of
Dominique Delande and Beno\^{\i}t Gr\'emaud
of the group ``Dynamique des syst\`emes coulombiens'' at 
the Laboratoire Kastler-Brossel.

\section*{Appendix: Extrema for the three-body Coulombian problem}\label{sec:appendixA}

For the three-body Coulombian problem, as can be seen from the second
sum in the right-hand side of~\eqref{eq:ElocNbodyCoulomb}, we must
find
\begin{equation}
  F_3^\mathrm{max}(a_1,a_2,a_3)\DEF\sup_{\mathrm{triangles}}[a_1\cos\theta_1+a_2\cos\theta_2+a_3\cos\theta_3]
\end{equation}
and 
\begin{equation}
  F_3^\mathrm{min}(a_1,a_2,a_3)\DEF\inf_{\mathrm{triangles}}[a_1\cos\theta_1+a_2\cos\theta_2+a_3\cos\theta_3]
\end{equation}
where the~$\theta$'s are the angles at the three vertices of the
triangle made of the three particles.  The $a$'s are some real
parameters that depend on the masses and the coupling constants.  Let
us define
\begin{equation}\label{def:f}
  f(a_1,a_2,a_3)\DEF\frac{1}{2}\left(\frac{a_1a_2}{a_3}+\frac{a_1a_3}{a_2}+\frac{a_2a_3}{a_1}\right)\;,
\end{equation}
then we have
\begin{equation}\label{eq:F3minmax}
   F_3^{\substack{\mathrm{max}\\\mathrm{min}}}(a_1,a_2,a_3)
  =\substack{\mathrm{max}\\{\mathrm{min}}}\{a_1+a_2-a_3,a_1-a_2+a_3,-a_1+a_2+a_3, f(a_1,a_2,a_3)\}
\end{equation}
$f$ is considered in the list~\eqref{eq:F3minmax} if and only
if the following three conditions are satisfied simultaneously
\begin{equation}\label{eq:ineqcos}
  \frac{1}{2}\left|\frac{a_2}{a_3}+\frac{a_3}{a_2}-\frac{a_2a_3}{a_1^2}\right|\leqslant1\;;
  \frac{1}{2}\left|\frac{a_1}{a_2}+\frac{a_2}{a_1}-\frac{a_1a_2}{a_3^2}\right|\leqslant1\;;
  \frac{1}{2}\left|\frac{a_3}{a_1}+\frac{a_1}{a_3}-\frac{a_1a_3}{a_2^2}\right|\leqslant1\;.
\end{equation}
Here is the proof: We will restrict the values of the~$\theta$'s to~$[0,\pi]$
 and the constraint $\theta_1+\theta_2+\theta_3=\pi$
is implemented by the Lagrange multiplier method. We are led to
extremalise the function $G_3(\theta_1,\theta_2,\theta_3)\DEF
a_1\cos\theta_1+a_2\cos\theta_2+a_3\cos\theta_3+\ell(\theta_1+\theta_2+\theta_3-\pi)$
for unconstrained~$(\theta_1,\theta_2,\theta_3)\in[0,\pi]^3$, $\ell$
being the Lagrange multiplier.  The three
conditions~$\partial_{\theta_i}G_3=0$ for~$i=1,2,3$ lead
to~$\ell=a_1\sin\theta_1=a_2\sin\theta_2=a_3\sin\theta_3$.  The
case~$\ell=0$ corresponds to the alignment of the three particles and
gives the three first values in the list~\eqref{eq:F3minmax}
corresponding to~$(\theta_1,\theta_2,\theta_3)=(0,0,\pi)$ and its
circular permutations.

Taking into account the constraint on the angles, we have
$\ell=a_3\sin\theta_3=a_3\sin\theta_1\cos\theta_2+a_3\sin\theta_2\cos\theta_1=\ell(a_3\cos\theta_2/a_1+a_3\cos\theta_1/a_2)$.
Therefore when~$\ell\neq0$, we find
$a_1\cos\theta_1+a_2\cos\theta_2=a_1a_2/a_3$ as well as the other
relations that are obtained by circular permutations of the
indices. From the decomposition
$F_3=\frac{1}{2}(a_1\cos\theta_1+a_2\cos\theta_2)+\frac{1}{2}(a_2\cos\theta_2+a_3\cos\theta_3)+\frac{1}{2}(a_1\cos\theta_1+a_3\cos\theta_3)$,
we obtain the value~\eqref{def:f} that must be considered
in~\eqref{eq:F3minmax} if and only if there exists some $\theta$'s
such that
\begin{equation}
  a_1\sin\theta_1=a_2\sin\theta_2=a_3\sin\theta_3 \quad\mathrm{and}\quad \theta_1+\theta_2+\theta_3=\pi\;.
\end{equation}
Solving these three equations leads to the values for the three
$|\cos\theta_i|$ that are precisely given by the left-hand sides of
the inequalities~\eqref{eq:ineqcos}.

\end{document}